\documentclass[onecolumn,showpacs,preprintnumbers,amsmath,amssymb,prb]{revtex4}
\usepackage{graphicx}% Include figure files
\usepackage{dcolumn}% Align table columns on decimal point
\usepackage{bm}% bold math
\usepackage{color}

\begin{document}

\preprint{APS/123-QED}

\title{Kondo effects and shot noise enhancement in a laterally coupled double quantum dot}% Force line breaks with \\

\author{Toshihiro Kubo$^{1,2}$}
 \email{kubo@will.brl.ntt.co.jp}
\author{Yasuhiro Tokura$^{1,2}$}
\author{Seigo Tarucha$^{1,3}$}%
 \affiliation{%
$^1$JST, ICORP, Quantum Spin Information Project, Atsugi-shi, Kanagawa 243-0198, Japan\\
$^2$NTT Basic Research Laboratories, NTT Corporation, Atsugi-shi, Kanagawa 243-0198, Japan\\
$^3$Department of Applied Physics, University of Tokyo, Hongo, Bunkyo-ku, Tokyo 113-8656, Japan
}%

\date{\today}% It is always \today, today,
             %  but any date may be explicitly specified

\begin{abstract}
The spin and orbital Kondo effects and the related shot noise for a laterally coupled double quantum dot are studied taking account of coherent indirect coupling via a reservoir. We calculate the linear conductance and shot noise for various charge states to distinguish between the spin and orbital Kondo effects. We find that a novel antiferromagnetic exchange coupling can be generated by the coherent indirect coupling, and it works to suppress the spin Kondo effect when each quantum dot holds just one electron. We also show that we can capture the feature of the pseudospin Kondo effect from the shot noise measurement.
\end{abstract}

% PACS, the Physics and Astronomy
                             % Classification Scheme.
%\keywords{Suggested keywords}%Use showkeys class option if keyword
                              %display desired
\maketitle

\section{Introduction\label{introduction}}
The Kondo effect was discovered many years ago in metals with dilute magnetic impurities and has long been studied as one of the most important many-body correlations in condensed matter physics \cite{kondo,kondo2}. We have obtained the physical understanding in equilibrium Kondo systems using the powerful methods such as exact solution and numerical renormalization group (RG) approach \cite{kondo3,kondo4} More recently it has been predicted that the Kondo effect occurs in semiconductor quantum dots (QDs) \cite{qd1,qd2}, and indeed, it has been observed in transport measurements for various kinds of QDs \cite{qd5}. In a single QD system, the Kondo effect gives rise to the enhancement of the conductance, and the conductance reaches the value of $2e^2/h$ at the unitary limit\cite{unitary1,unitary2}. Since then the Kondo effect in QDs has been attracting a lot of new interests associated with extended degrees of freedom, such as tunnel coupling to reservoirs, the number of trapped electrons in a QD, and the number of Kondo channels. By tuning these parameters, various aspects of the Kondo effect have been revealed including enhancement induced by state degeneracy \cite{qdexp1}, the unitary limit \cite{qdexp3}, and the nonequilibrium Kondo effect \cite{qdexp4}. Therefore, QDs are regarded as artificial Kondo systems, in which various theoretical approximations can be tested to acquire a better understanding of strongly correlated electron systems. In particular, the nonequilibrium Kondo problem is not yet solved completely despite the large number of theoretical studies. The nonequilibrium magnetization of the QD was revisited using the Schwinger-Keldysh perturbation formalism\cite{non1}. When the large bias voltage or a magnetic field is applied, the transport through the QD was studied by the perturbative RG approach\cite{non2,non3,non4} (so-called poor man's scaling developed by Anderson\cite{poor}). Using the real-time perturbation theory in Schwinger-Keldysh formulation, the universal properties that the perturbative series of any average in the steady state satisfies the equilibrium Callan-Symanzik equations\cite{non5}. By the real-time RG in frequency space, the nonequilibrium anisotropic Kondo model was examined in the weak coupling regime, where the maximum of bias voltage and magnetic field is larger than the Kondo temperature\cite{non6}. In the framework of the same approach, the dynamical spin-spin correlation function was calculated in nonequilibrium Kondo systems describing spin and/or orbital fluctuations\cite{non7}. Using the generalized flow equation approach to include a magnetic field similar to the real-time RG performed by Schoeller \textit{et al.}, the spin-spin correlation function, the $T$-matrix, and the magnetization were calculated as a function of applied magnetic field\cite{non8}. By a nonequilibrium RG method, the real-time evolution of spin and current in the anisotropic Kondo model were investigated at a finite magnetic field and bias voltage\cite{non9}.

Recently, in not only the single QD but also the double quantum dot (DQD) systems, the Kondo effects have been studied. In particular, the interplay between the Kondo effect and inter-dot correlation was discussed \cite{sok1,sok2,sok3,sok4}. It is theoretically predicted that the two-channel Kondo model realized in a DQD system exhibits a non-Fermi liquid quantum critical point\cite{nfl}. Such a two-channel Kondo problem was experimentally investigated\cite{nfl2}. Moreover, the Kondo problem is more intriguing in DQDs than in single QDs because of the competition between the spin Kondo effect and the orbital (pseudospin) Kondo effect \cite{dqd1,dqd2,dqd4,ok1,final}. In a DQD, the pseudospin state is represented as a state with an electron in either of two capacitively coupled QDs but separately contacted to a pair of reserviors (see Fig. \ref{fig1}(a)). It has been predicted that the $SU(4)$ Kondo effect will provide a novel feature for a highly symmetric DQD configuration \cite{dqd4,eto}. However, it is difficult to realize the $SU(4)$ condition experimentally since the intra-dot Coulomb interaction is usually larger than the inter-dot Coulomb interaction. The pseudospin Kondo effect is only defined in DQDs, and has recently been confirmed experimentally, but not in reference to the interplay with the spin Kondo effect \cite{dqdps}. In contrast with an ideal DQD as shown in Fig. \ref{fig1}(a), most experiments are performed for DQDs with an integrated reservoir (see Fig. \ref{fig1}(b)). In such DQDs, the pseudospin-dependent linewidth function is induced by the coherent indirect coupling via the integrated reservoir \cite{kubo2}. The effect of the indirect coupling on the spin Kondo effect in DQDs with integrated reservoirs have been discussed only where the indirect coupling is at its maximum value as shown in Fig. \ref{fig1}(c) \cite{dqd6,dqd7,konik}. However, most of the actual experimental conditions correspond to an intermediate condition (for example \cite{dqd5}), and so it is important to study the effects of indirect coupling on the Kondo effect. Moreover, theoretical studies often focus on a situation where two QDs are energetically aligned. The pseudospin Kondo effect strongly depends on the charge states in the DQD. Therefore, it is useful to employ the entire charge state diagram to capture the signature of the pseudospin Kondo effect.

In this paper, we investigate the effects of coherent indirect coupling via a reservoir on the Kondo effects in a laterally coupled DQD. We employ the finite Coulomb interaction slave-boson mean-field theory (SBMFT) \cite{sbmft} using the nonequilibrium Green's function method. This approach allows us to take account of the coherence between two QDs nonperturbatively. To characterize the indirect coupling, we introduce a coherent indirect coupling parameter $\alpha$ \cite{raikh,kubo1}. For finite $\alpha$, the pseudospin Kondo effect is suppressed since the linewidth function depends on the pseudospin and the $SU(2)$ symmetry is broken \cite{kubo2}. Here we newly find that the coherent indirect coupling leads to novel antiferromagnetic kinetic exchange coupling between two local spins in QDs via the reservoir. This kinetic exchange coupling via the reservoir competes with the Kondo exchange coupling, and hence the spin Kondo effect is suppressed when each QD holds just one electron. Such a phenomenon can occur in parallel but not series coupled DQDs. Then, we examine the shot noise to devise a new approach for characterizing the pseudospin Kondo effect. To distinguish between the spin Kondo effect and pseudospin Kondo effect can be difficult in standard conductance measurements. We find that the shot noise experiment can provide a clear contrast between them. The shot noise has recently been discussed extensively in relation to charge fluctuations in mesoscopic systems \cite{noise}. The current noise $S(\omega)$ is defined by a Fourier transform of $S(t,t')=\langle \{\delta I(t),\delta I(t')\} \rangle$, where $\delta I(t)\equiv I(t)-\langle I(t)\rangle$ is the amount by which the current deviates from its average value. The equilibrium zero-frequency current noise $S(0)$ cannot carry additional information beyond the conductance. In contrast, the \textit{nonequilibrium} zero-frequency shot noise can provide information on charge fluctuations, which is not accessible in conventional transport measurements. The source-drain bias voltage dependence of the shot noise through a QD in the spin Kondo regime has been studied theoretically \cite{qdsn1}. Here, the pseudospin Kondo effect is generally promoted by the charge fluctuation, so we examine the shot noise in the charge stability diagram, and find that it is strongly enhanced in the pseudospin Kondo regime.

This paper is organized as follows. In Sec. \ref{model}, a standard tunneling Hamiltonian is employed to describe a laterally coupled DQD. We introduce the notion of the coherent indirect coupling for the source reservoir. We provide the expressions of the linear conductance and  the zero-frequency shot noise at zero temperature using the nonequilibrium Green's function method.  We discuss the numerical results for the linear conductance and zero-frequency shot noise at zero temperature in Sec. \ref{result}. In particular, we derive the new antiferromagnetic kinetic exchange coupling induced by a coherent indirect coupling via the reservoir. We show that the spin-spin correlation is antiferromagnetic. All our results are summarized in Sec. \ref{conclusion}. In Appendix \ref{exchange}, we provide the detailed derivation of the effective spin-spin Hamiltonian with an antiferromagnetic kinetic exchange coupling induced by a coherent indirect coupling using the 4th-order Rayleigh-Schr\"{o}dinger degenerate perturbation theory.

\section{Model and formulation\label{model}}
\begin{figure}
\includegraphics[scale=0.4]{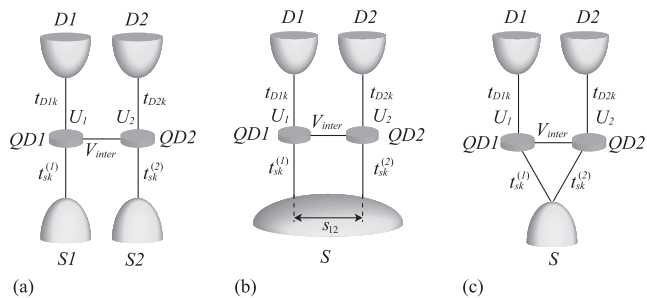}% Here is how to import Epseudospin art
\caption{\label{fig1} Schematic diagrams of laterally coupled DQDs with a separated drain reservoir. $s_{12}$ is the minimum distance that electrons propagate in the source reservoir. (a) The source and drain reservoirs are both completely separated, namely there is no coupling between two QDs via the reservoirs. This situation corresponds to $s_{12}\to\infty$. (b) The source reservoir is common and the drain reservoir is separated. (c) There is maximal coherence between two QDs via the reservoirs. This condition corresponds to $s_{12}=0$.}
\end{figure}

We consider a DQD tunnel coupled to one common source reservoir and two drain reservoirs as shown in Fig. \ref{fig1}(b). We assume only a single energy level for each QD. The Hamiltonian represents the sum of the following terms: $H=H_R+H_{DQD}+H_T$. The Hamiltonian of the Fermi liquid reservoirs is
\begin{equation}
H_R=\sum_{\nu\in\{S,D1,D2\}}\sum_k\sum_{\sigma\in\{\uparrow,\downarrow\}}\epsilon_{\nu k}{a_{\nu k\sigma}}^{\dagger}a_{\nu k\sigma},
\end{equation}
where $\epsilon_{\nu k}$ is the electron energy with wave number $k$ in the reservoir $\nu$ and the operator $a_{\nu k\sigma}$ (${a_{\nu k\sigma}}^{\dagger}$) annihilates (creates) an electron with spin $\sigma$ in the reservoirs. $H_{DQD}$ describes an isolated DQD,
\begin{eqnarray}
H_{DQD}=\sum_{i=1}^2\sum_{\sigma\in\{\uparrow,\downarrow\}}\epsilon_in_{i\sigma}+\sum_{i=1}^2U_in_{i\uparrow}n_{i\downarrow}+V_{inter}\sum_{\sigma\in\{\uparrow,\downarrow\}}\sum_{\sigma'\in\{\uparrow,\downarrow\}}n_{1\sigma}n_{2\sigma'},
\end{eqnarray}
where $\epsilon_i$ is the energy level of the $i$th QD, $U_i$ is the on-site Coulomb interaction in the $i$th QD, and $V_{inter}$ is the inter-dot Coulomb interaction. Here the following notations are introduced: $c_{i\sigma}$ (${c_{i\sigma}}^{\dagger}$) is an annihilation (creation) operator of an electron in the $i$th QD with spin $\sigma$ and $n_{i\sigma}={c_{i\sigma}}^{\dagger}c_{i\sigma}$ is its number operator. The tunneling Hamiltonian between the QDs and source and drain reservoirs is given by
\begin{equation}
H_T=\sum_k\sum_{i=1}^2\sum_{\sigma\in\{\uparrow,\downarrow\}}\left[t_{Sk}^{(i)}{a_{Sk\sigma}}^{\dagger}c_{i\sigma}+t_{Dik}{a_{Dik\sigma}}^{\dagger}c_{i\sigma}+\mbox{H.c.} \right].\label{tunnel}
\end{equation}
We take account of the propagation process of electrons in the source reservoir. This leads to coherent indirect coupling between two QDs via the source reservoir \cite{raikh,kubo1}, which is characterized by a parameter $\alpha$. The linewidth function matrices are then given by
\begin{equation}
\bm{\Gamma}_{\sigma}^S=\Gamma_S\left(
  \begin{array}{cc}
    1   &  \alpha  \\
    \alpha   &  1  \\
  \end{array}
\right)\ ,\ \bm{\Gamma}_{\sigma}^{D1}=\Gamma_D\left(
  \begin{array}{cc}
    1   &  0  \\
    0   &  0  \\
  \end{array}
\right)\ ,\ \bm{\Gamma}_{\sigma}^{D2}=\Gamma_D\left(
  \begin{array}{cc}
    0   &  0  \\
    0   &  1  \\
  \end{array}
\right),
\end{equation}
where the boldface notation indicates a matrix whose basis is the localized state in each QD. Here we assume that $|t_{Sk}^{(1)}|^2=|t_{Sk}^{(2)}|^2\equiv|t_{Sk}|^2$ and $|t_{D1k}|^2=|t_{D2k}|^2\equiv|t_{Dk}|^2$. The linewidth function is defined by $\Gamma_{\nu}(\epsilon)\equiv(2\pi/\hbar)\sum_k|t_{\nu k}|^2\delta(\epsilon-\epsilon_{\nu k})$, and we neglected its energy dependence in the wide-band limit, namely $\Gamma_{\nu}(\epsilon)=\Gamma_{\nu}$. $\alpha$ is a function of the propagation length $s_{12}$ of the electrons in the reservoir \cite{kubo1}, and in general, $|\alpha|\le 1$. The condition $s_{12}=0$ is equivalent to $\alpha=1$ \cite{alpha}. The importance of the sign of the coherent indirect coupling parameter was pointed out by S. A. Gurvitz \cite{sign}. The wave number dependence of the tunneling amplitude $t_{\nu k}^{(j)}$ is usually neglected in the theoretical treatment. However, in case the two QDs are indirectly coupled via the source reservoir as assumed here, the wave number dependence of the tunneling amplitude plays an important role in generating an indirect hopping between the QDs. As explained later, such an indirect hopping process causes an antiferromagnetic kinetic exchange coupling. The mechanism is similar to that by a direct inter-dot coupling mechanism \cite{sok3}, however, for the indirect inter-dot coupling, the exchange coupling constant includes information of coherence in the source reservoir.

We use the finite Coulomb interaction SBMFT \cite{sbmft} to investigate the linear conductance and shot noise through DQDs. In this approach, the slave-boson operators are replaced by nonfluctuating average values, leading to a noninteracting resonant tunneling model, whose $28$ unknown parameters have to be determined self-consistently. The result obtained with this method agrees fairly well with a numerical Lanczos calculation and a numerical renormalization group calculation for a tunnel-coupled DQD \cite{lanczos,sok3,dong}.

The tunneling current through a DQD can be expressed in terms of the transmission matrix \cite{current},
\begin{equation}
I=\frac{e}{h}\sum_{i=1}^2\sum_{\sigma\in\{\uparrow,\downarrow\}}\int d\epsilon[f_S(\epsilon)-f_{Di}(\epsilon)]\mbox{Tr}\left\{\bm{T}_{i\sigma}(\epsilon) \right\}.
\end{equation}
Here the transmission matrix is defined as $\bm{T}_{i\sigma}(\epsilon)=\bm{G}_{\sigma}^r(\epsilon)\bm{\Gamma}_{\sigma}^S\bm{G}_{\sigma}^a(\epsilon)\bm{\Gamma}_{\sigma}^{Di}$ using the retarded (advanced) Green's function $\bm{G}_{\sigma}^r(\epsilon)$ ($\bm{G}_{\sigma}^a(\epsilon)$) of the DQD, and $f_{\nu}(\epsilon)=1/(1+e^{(\epsilon-\mu_{\nu})/k_BT})$ is the Fermi-Dirac distribution function in the reservoir $\nu$ at temperature $T$. Within the framework of the finite Coulomb interaction SBMFT, the retarded Green's function is given by
\begin{eqnarray}
\bm{G}_{\sigma}^r(\epsilon)&=&\left(
  \begin{array}{cc}
    \frac{\epsilon-\tilde{\epsilon}_1}{\hbar}+\frac{i}{2}\tilde{\Gamma}_{11,\sigma}   & \frac{i}{2}\tilde{\Gamma}_{12,\sigma}   \\
    \frac{i}{2}\tilde{\Gamma}_{21,\sigma}   &  \frac{\epsilon-\tilde{\epsilon}_2}{\hbar}+\frac{i}{2}\tilde{\Gamma}_{22,\sigma}  \\
  \end{array}
\right)^{-1},
\end{eqnarray}
where $\tilde{\epsilon}_i$ and $\tilde{\Gamma}_{ij,\sigma}$ are the renormalized energy level of the $i$th QD and the $(i,j)$ matrix element of the linewidth function matrix for spin $\sigma$. Such renormalizations indicate the Coulomb interaction effects. The advanced Green's function is obtained from the retarded Green's function: $\bm{G}_{\sigma}^a(\epsilon)=[\bm{G}_{\sigma}^r(\epsilon)]^{\dagger}$. The source and drain reservoirs have chemical potentials $\mu_S=\mu+eV_{SD}/2$ and $\mu_{Di}=\mu-eV_{SD}/2$ with the source-drain bias voltage $V_{SD}$, and $\mu=0$ is the origin of the energy. Here we assume that the two drain reservoirs have the same chemical potential. In the following, we focus on the zero temperature condition. Then, the linear conductance through the $i$th QD is given by
\begin{equation}
G_i=\frac{e^2}{h}\sum_{\sigma\in\{\uparrow,\downarrow\}}T_{i\sigma},\label{conductance}
\end{equation}
where $T_{i\sigma}\equiv\mbox{Tr}\left\{\bm{T}_{i\sigma}(0) \right\}$ is the transmission probability of the conduction channel for spin $\sigma$ in the $i$th QD. Within the framework of the SBMFT, the zero-frequency shot noise is given by the Khlus-Lesovik formula \cite{noise2,noise3},
\begin{eqnarray}
S(0)&=&\frac{e^2}{\pi}\sum_{i=1}^2\sum_{\sigma\in\{\uparrow,\downarrow\}}\int_{-eV_{SD}/2}^{eV_{SD}/2}\frac{d\epsilon}{\hbar}\mbox{Tr}\left\{\bm{T}_{i\sigma}(\epsilon)\left[1-\bm{T}_{i\sigma}(\epsilon) \right] \right\}\nonumber\\
&=&\frac{e^2}{\pi}\sum_{i=1}^2\sum_{\sigma\in\{\uparrow,\downarrow\}}\int_{-eV_{SD}/2}^{eV_{SD}/2}\frac{d\epsilon}{\hbar}T_{i\sigma}(\epsilon)[1-T_{i\sigma}(\epsilon)],\label{shot}
\end{eqnarray}
where $T_{i\sigma}(\epsilon)=\mbox{Tr}\left\{\bm{T}_{i\sigma}(\epsilon) \right\}$. In our problem, although the transmission matrix has finite off-diagonal elements for $\alpha\neq0$, the zero-frequency shot noise can be expressed as Eq. (\ref{shot}) in terms of the simple summation of $T_{i\sigma}(\epsilon)[1-T_{i\sigma}(\epsilon)]$ for each conduction mode since the drain reservoirs are separated and there is no indirect coupling.

\section{Theoretical results\label{result}}
\subsection{Linear transport}
\begin{figure}
\includegraphics[scale=0.3]{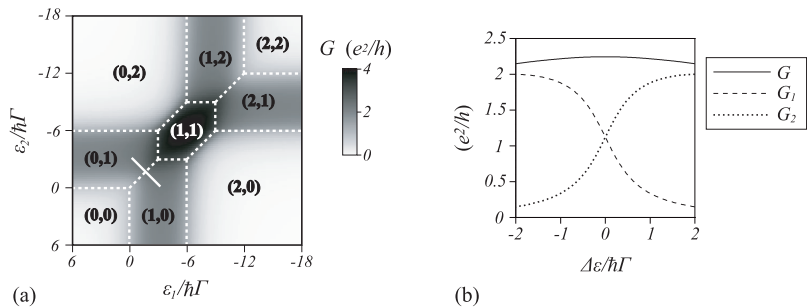}% Here is how to import Epseudospin art
\caption{\label{fig2} Total linear conductance $G$ for $\alpha=0$ and $U_1/\hbar\Gamma=U_2/\hbar\Gamma=2V_{inter}/\hbar\Gamma=6$. (a) The charge configuration is shown as $(N_1,N_2)$. The white dotted line indicates the charge degeneracy line schematically. (b) $\Delta\epsilon$ dependence of the linear conductance along the white line in (a). The broken, dotted, and solid lines indicate the conductance $G_1$, $G_2$, and the total conductance $G$, respectively.}
\end{figure}

In the following discussions, we assume that $U_1/\hbar\Gamma=U_2/\hbar\Gamma\equiv U/\hbar\Gamma=2V_{inter}/\hbar\Gamma=6$, and $\Gamma_S=\Gamma_D=\Gamma$ as a typical example, and to show the charge configurations, we introduce the notation $(N_1,N_2)$, where $N_i$ is the population of the $i$th QD. First, we consider the situation without coherent indirect coupling, namely $\alpha=0$, shown in Fig. \ref{fig1}(a). The total linear conductance $G=G_1+G_2$ is shown in Fig. \ref{fig2}(a) as a function of $\epsilon_1$ and $\epsilon_2$ (charge stability diagram). The conductance is suppressed owing to the Coulomb blockade in the $(0,0)$, $(2,0)$, $(0,2)$, and $(2,2)$ regimes. In the $(1,0)$, $(0,1)$, $(2,1)$, and $(1,2)$ regimes, $G\simeq 2e^2/h$ since the conductance is enhanced as a result of the spin Kondo effects. In the $(1,1)$ regime, we have the double spin Kondo effect, namely spin Kondo effects in each QD, and the conductance value reaches $4e^2/h$. Without depending on the ratio between $U$ and $V_{inter}$, the linear conductance can reach $4e^2/h$ at $\epsilon_1/\hbar\Gamma=\epsilon_2/\hbar\Gamma=-(U/2+V_{inter})/\hbar\Gamma$, namely the center of the $(1,1)$ region \cite{oguchi}. In Fig. \ref{fig2}(b), we plot the energy offset $\Delta\epsilon(\equiv\epsilon_1-\epsilon_2)$ dependence of the linear conductance along the white line in Fig. \ref{fig2}(a). $\Delta\epsilon=0$ corresponds to $\epsilon_1=\epsilon_2=-V_{inter}/2$. For the spinless electrons, the linear conductance cannot exceed $2e^2/h$ in the pseudospin Kondo regime, namely the $(1,0)-(0,1)$, $(2,0)-(1,1)$, $(1,1)-(0,2)$, and $(2,1)-(1,2)$ boundaries. However, when the spin and pseudospin degrees of freedom are entangled, $G$ exceeds $2e^2/h$ as shown in Fig. \ref{fig2}(b). For a large $\Delta\epsilon$, $G$ approaches $2e^2/h$ since the situation becomes equivalent to that of the spin Kondo regime in a single QD. These results are qualitatively consistent with those obtained with the numerical renormalization group method \cite{eto}.

\begin{figure}
\includegraphics[scale=0.3]{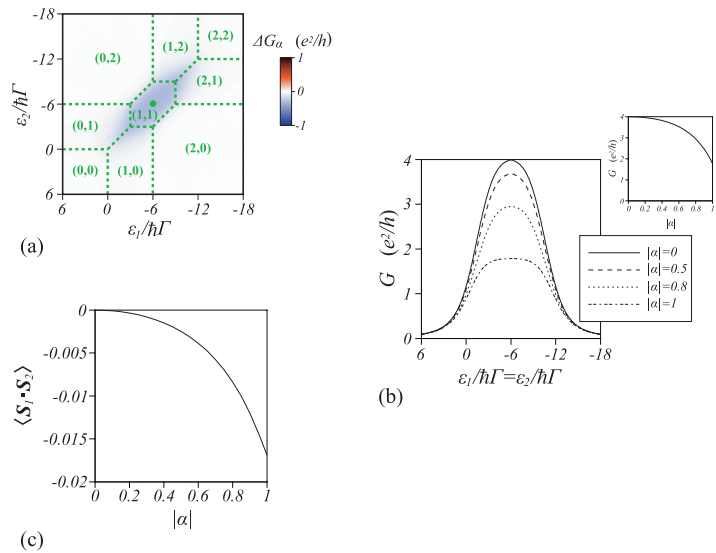}% Here is how to import Epseudospin art
\caption{\label{fig3} Reduction of the linear conductance caused by the coherent indirect coupling and the spin-spin correlation function. (a) $\Delta G_{\alpha}$ for $\alpha=0.5$. (b) $G$ for $\epsilon_1=\epsilon_2$. The solid, broken, dotted, and dash-dotted line lines indicate $\alpha=0$, $|\alpha|=0.5$, $|\alpha|=0.8$, and $|\alpha|=1$, respectively. Inset: $|\alpha|$ dependence of the linear conductane at $\epsilon_1/\hbar\Gamma=\epsilon_2/\hbar\Gamma=-6$ indicated by the green circle in (a). (c) $|\alpha|$ dependence of the spin-spin correlation function at $\epsilon_1/\hbar\Gamma=\epsilon_2/\hbar\Gamma=-6$ indicated by the green circle in (a).}
\end{figure}

Next, we consider the effect of $\alpha$. In Fig. \ref{fig3}(a), we show the conductance difference $\Delta G_{\alpha}$ between $G$ of $\alpha=0.5$ and $G$ of $\alpha=0$. From Fig. \ref{fig3}(a), we find that the linear conductance decreases only in the $(1,1)$ charge configuration. In this $(1,1)$ charge configuration, the coherent indirect coupling gives rise to antiferromagnetic kinetic exchange coupling as follows: We consider the tunneling Hamiltonian (\ref{tunnel}) as a perturbation, and we calculate the effective spin-spin interaction Hamiltonian using the 4th-order Rayleigh-Schr\"{o}dinger degenerate perturbation theory, namely the effective Hamiltonian is given as $H_{eff}^{\alpha}=H_T\frac{1}{E-H_0}H_T\frac{1}{E-H_0}H_T\frac{1}{E-H_0}H_T$, where $H_0\equiv H_R+H_{DQD}$ is the unperturbed Hamiltonian. As a result, we obtain the following effective spin-spin interaction Hamiltonian: $H_{eff}^{\alpha}\simeq J_{\alpha}\bm{S}_1\cdot\bm{S}_2$ with
\begin{equation}
J_{\alpha}=16\epsilon_F\left(\frac{\alpha\hbar\Gamma}{\pi U} \right)^2,\label{ex}
\end{equation}
where $\bm{S}_i$ is the spin operator of the $i$th QD and $\epsilon_F$ is the Fermi energy. Here we consider the possibility to observe this exchange coupling experimentally. To observe the Kondo effect, we usually use the QD systems in the strong coupling regime, namely large $\Gamma$, since the Kondo temperature becomes higher. Thus, we expect to be experimentally possible to verify the antiferromagnetic kinetic exchange interaction induced by the coherent indirect coupling in the strong coupling QD systems since the factor $\hbar\Gamma/U$ in Eq. (\ref{ex}) is not small in such systems. We provide the detailed derivation of this antiferromagnetic kinetic exchange interaction in Appendix \ref{exchange}. This kinetic exchange coupling competes with the Kondo exchange coupling. Therefore, in the $(1,1)$ regime, the spin Kondo effect is suppressed with the increase in $|\alpha|$ and hence the conductance decreases as shown in Fig. \ref{fig3}(b). This suppression is independent of the sign of $\alpha$. In inset of Fig. \ref{fig3}(b), we plot the $|\alpha|$ dependence of the conductance when $\epsilon_1/\hbar\Gamma=\epsilon_2/\hbar\Gamma=-6$ as indicated by green circle in Fig. \ref{fig3}(a). The linear conductance decreases monotonically with increasing $|\alpha|$. Similarly, we show the spin-spin correlation function $\langle \bm{S}_1\cdot\bm{S}_2 \rangle$ in Fig. \ref{fig3} (c). We evaluate the spin-spin correlation function $\langle \bm{S}_1\cdot\bm{S}_2 \rangle$ using the nonequilibrium Green's functions as follows:
\begin{eqnarray}
\langle \bm{S}_1\cdot\bm{S}_2 \rangle=\frac{3}{8\pi^2}\int d\omega\int\frac{d\epsilon}{\hbar}G_{21,\sigma}^{-+}(\epsilon)G_{12,\sigma}^{+-}(\epsilon+\hbar\omega),
\end{eqnarray}
where $G_{ij,\sigma}^{-+}(\epsilon)$ and $G_{ij,\sigma}^{+-}(\epsilon)$ are the $(i,j)$ matrix element of the lesser and greater Green's functions for spin $\sigma$. These can be obtained from the retarded and advanced Green's functions using the Keldysh equation \cite{keldysh} as follows
\begin{eqnarray}
\bm{G}_{\sigma}^{-+}(\epsilon)&=&i\sum_{\nu\in\{S,D1,D2\}}f_{\nu}(\epsilon)\bm{G}_{\sigma}^r(\epsilon)\bm{\Gamma}_{\sigma}^{\nu}\bm{G}_{\sigma}^a(\epsilon),\\
\bm{G}_{\sigma}^{+-}(\epsilon)&=&-i\sum_{\nu\in\{S,D1,D2\}}[1-f_{\nu}(\epsilon)]\bm{G}_{\sigma}^r(\epsilon)\bm{\Gamma}_{\sigma}^{\nu}\bm{G}_{\sigma}^a(\epsilon).
\end{eqnarray}
When $|\alpha|$ increases, $\langle \bm{S}_1\cdot\bm{S}_2 \rangle$ increases negatively. This means that the antiferromagnetic kinetic exchange coupling becomes dominating as $|\alpha|$ increases.

The Ruderman-Kittel-Kasuya-Yosida (RKKY) interaction is well known as an indirect exchange interaction between two local spins \cite{rkky1,rkky2,rkky3}. In the RKKY interaction the exchange coupling becomes weak with changing the sign between positive and negative, and therefore changing the magnetic character between ferromagnetic and antiferromagnetic as the two local spins become separated from each other. The RKKY interaction in semiconductor QD systems has been studied both theoretically and experimentally \cite{rkky4,rkky5}. Particularly when $\alpha=1$ for both source and drain reservoirs with $\Delta\epsilon\neq0$, Konik discussed the RKKY-Kondo like effect in a similar type DQD \cite{konik}. In contrast we concentrate on the competition between the Kondo exchnage and $J_{\alpha}$ when $\Delta\epsilon=0$ and $\alpha=0$ for the drain reservoir. If we investigate this competition when $\Delta\epsilon=0$ and $\alpha=1$ for both the source and drain reservoirs, we expect the single channel Kondo effect (the exchange coupling caused by the coherent indirect coupling vanishes as shown in Appendix \ref{spin-spin}) since there is only a single conduction mode in such a situation, namely one of the two orbital channels is in a dark state \cite{kubo1}. Here, although we considered the effect of the integrated reservoir only for the source, we can expect stronger suppression of the spin Kondo effect in the $(1,1)$ regime when both the source and drain reservoirs are integrated with $0<|\alpha|\neq1$. It is noted that the kinetic exchange coupling induced by a coherent indirect coupling is different from the RKKY exchange coupling. The main difference between these two exchange interactions is where the dependence of the inter-dot distance is included. In our exchange interaction the coherent indirect coupling parameter $\alpha$ is a decision factor, and the interaction strength is proportional to $|\alpha|^2$, and the magnetic character is always antiferromagnetic. Note that $\alpha$ becomes small and changes the sign with increasing distance between the two local spins. In the RKKY interaction, although the wave number dependence of the response function is considered, the wave number dependence of the tunneling amplitude $t_{Sk}^{(i)}$ is neglected, just like the case for an impurity as a point scatterer, to account for the oscillatory behavior of the exchange coupling with the distance between the impurities. However, it is very important to take account of the wave number dependence of the tunneling amplitude in DQD systems since the wave function of electron confined in QDs relatively spreads, and the tunnel couplings are highly anisotropic. Thus, in the present problem, we believe that it is preferable to discuss our exchange interaction in terms of the coherent indirect coupling than the RKKY exchange interaction.

\subsection{Shot noise}
\begin{figure}
\includegraphics[scale=0.3]{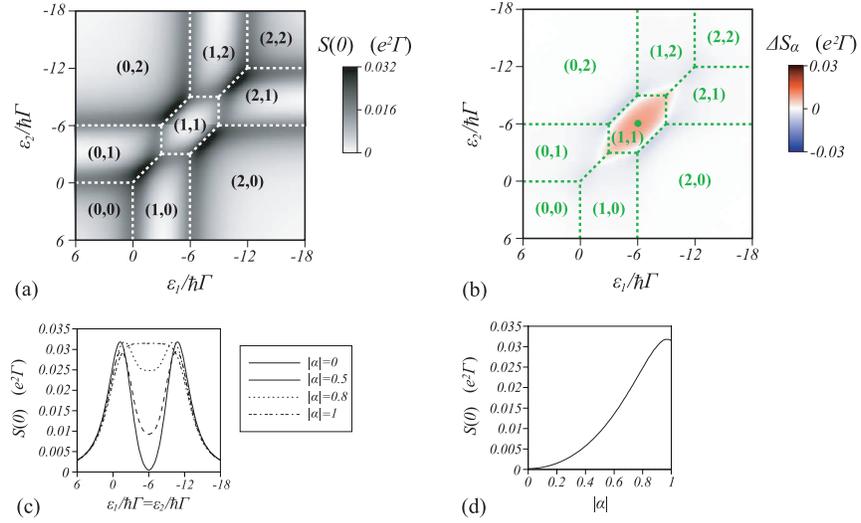}% Here is how to import Epseudospin art
\caption{\label{fig4} Shot noise $S(0)$ and shot noise difference $\Delta S_{\alpha}$ for $U_1/\hbar\Gamma=U_2/\hbar\Gamma=6$, $V_{inter}/\hbar\Gamma=3$, and $eV_{SD}/\hbar\Gamma=0.1$. The charge configuration is shown as $(N_1,N_2)$. (a) $S(0)$ for $\alpha=0$. (b) $\Delta S_{\alpha}$ for $|\alpha|=0.5$. (c) $S(0)$ for $\epsilon_1=\epsilon_2$. The solid, broken, dotted, and dash-dotted line lines indicate $\alpha=0$, $|\alpha|=0.5$, $|\alpha|=0.8$, and $|\alpha|=1$, respectively. (d) $|\alpha|$ dependence of the zero-frequency shot noise at $\epsilon_1/\hbar\Gamma=\epsilon_2/\hbar\Gamma=-6$ indicated by the green circle in (b).}
\end{figure}

Interplay or competition between the spin and pseudospin Kondo effects can appear in the linear transport characteristic as shown in Fig. \ref{fig2}(b). However, it is still difficult to distinguish their contributions. This is particularly the case in experiments, because the spin Kondo conductance observed for single QDs is usually less than $2e^2/h$. To capture the feature of the pseudospin Kondo effect, which originates with the charge fluctuation, we investigate the shot noise, which provides information on charge fluctuations. In the following, we focus on the condition where $eV_{SD}/\hbar\Gamma=0.1$. First, we consider the situation without coherent indirect coupling. The zero-frequency shot noise is shown in Fig. \ref{fig4}(a). In the $(0,0)$, $(2,0)$, $(0,2)$, and $(2,2)$ regimes, the shot noise is strongly suppressed because of the Coulomb blockade. In the $(1,0)$, $(0,1)$, $(1,1)$, $(2,1)$, and $(1,2)$ regimes, the shot noise is also strongly suppressed since a perfect transmission is realized by the spin Kondo effect. By contrast, the zero-frequency shot noise is enhanced at the Coulomb peaks owing to the maximum charge fluctuations in one of the two QDs. As an example, we consider the $(0,0)-(1,0)$ boundary. In this situation, the transmission probabilities $T_{1\uparrow}$ and $T_{1\downarrow}$ of two conduction channels for the up and down spins in QD1, respectively, are $T_{1\uparrow}=T_{1\downarrow}=1/2$, and thus the shot noise becomes large. Moreover, the zero-frequency shot noise is enhanced in the pseudospin Kondo regimes, because the charge fluctuation is maximal, as shown in Fig. \ref{fig4}(a). In the pseudospin Kondo regimes, there can be four conduction channels, for example, for the $(1,0)-(0,1)$ boundary, $T_{1\uparrow}=T_{1\downarrow}=T_{2\uparrow}=T_{2\downarrow}\simeq 1/2$. As a result, the shot noise in the pseudospin Kondo regimes is about double that at the Coulomb peaks. Therefore, the shot noise in the charge stability diagram is maximal in the pseudospin Kondo regime, and the signature can be easily captured experimentally. It should be noted that the shot noise enhancement discussed here cannot be obtained in calculations of the mean-field level such as the Hartree-Fock approximation, and thus the many-body correlation is essential.

Next, we discuss the effects of the coherent indirect coupling on the shot noise. First, in Fig. \ref{fig4}(b), we show the shot noise difference $\Delta S_{\alpha}$ between $S(0)$ of $|\alpha|=0.5$ and $S(0)$ of $\alpha=0$. We found that the spin Kondo effects are suppressed with $|\alpha|$ in the $(1,1)$ regime. In this regime, the transmission probabilities of all the conduction channels become smaller than 1 due to the kinetic antiferromagnetic exchange coupling induced by the coherent indirect coupling. As a result, the shot noise becomes large. We plot the QD energy dependence of the shot noise as shown in Fig. \ref{fig4}(c). When $|\alpha|$ increases, the shot noise is mainly affected in the $(1,1)$ regime. In Fig. \ref{fig4}(d), we plot the $|\alpha|$ dependence of the shot noise when $\epsilon_1/\hbar\Gamma=\epsilon_2/\hbar\Gamma=-6$ indicated by the green circle in Fig. \ref{fig4}(b). The value of transmission probabilities for all conduction modes are the same since we consider the condition when the two QD energies align. As shown in Fig. \ref{fig3}(c), the value of the transmission probability for each conduction mode are approximately equal to $1/2$ at $|\alpha|\sim 0.97$ under low bias voltage since the linear conductance is proportional to the transmission probability (see Eq. (\ref{conductance})). Therefore, from Eq. (\ref{shot}), the zero-frequency shot noise becomes maximal at $|\alpha|\sim 0.97$.

\section{Conclusions\label{conclusion}}
To conclude, we have studied the effects of inter-dot coherent indirect coupling via the reservoir on the Kondo effect and shot noise in a laterally coupled DQD using the finite-Coulomb interaction SBMFT to demonstrate the significance of many-body correlations. In particular, we found that the coherent indirect coupling gives rise to antiferromagnetic kinetic exchange coupling using the 4th-order Rayleigh-Schr\"{o}dinger perturbation theory. Thus the spin Kondo effect is suppressed in the $(1,1)$ regime. To support that the new exchange coupling is antiferromagnetic, we estimate the spin-spin correlation function. The spin-spin correlation function increases negatively as the coherent indirect coupling parameter increases. We discussed the difference between the RKKY exchange coupling and the new antiferromagnetic exchange coupling induced by the coherent indirect coupling. Moreover, we suggested that shot noise measurement is more appropriate than conductance measurement for capturing the signature of the pseudospin Kondo effect, because the shot noise is strongly enhanced in the pseudospin Kondo regime.

\begin{acknowledgments}
We thank S. A. Gurvitz, A. Oguri, T. Aono, S. Sasaki, H. Oguchi, S. Amaha, T. Hatano, S. Teraoka, and Y.-S. Shin for useful discussions. Part of this work is supported financially by JSPS Grant-in-Aid for Scientific Research S (No. 19104007), MEXT Grant-in-Aid for Scientific Research on Innovative Areas (21102003), Funding Program for World-Leading Innovative R\&D Science and Technology (FIRST), and DARPA QuEST grant HR0011-09-1-0007.
\end{acknowledgments}

\appendix

\section{\label{exchange}Derivation of antiferromagnetic kinetic exchange interaction by coherent indirect coupling}
Here we show the detailed derivation of the antiferromagnetic kinetic exchange interaction induced by the coherent indirect coupling as discussed in Sec. \ref{result}. Starting from the state ${d_{1\uparrow}}^{\dagger}{d_{2\downarrow}}^{\dagger}|F\rangle$, where the state $|F\rangle$ corresponds to the Fermi seas of conduction electrons in the source reservoir $S$ with empty DQD, we consider the tunneling Hamiltonian as a perturbation and derive the effective spin-spin interaction Hamiltonian using the 4th-order Rayleigh-Schr\"{o}dinger degenerate perturbation theory. Then, we consider the following process:
\begin{equation}
H_{eff}^{\alpha}{d_{1\uparrow}}^{\dagger}{d_{2\downarrow}}^{\dagger}|F\rangle=H_T\frac{1}{E-H_0}H_T\frac{1}{E-H_0}H_T\frac{1}{E-H_0}H_T{d_{1\uparrow}}^{\dagger}{d_{2\downarrow}}^{\dagger}|F\rangle,
\end{equation}
where
\begin{equation}
H_0\equiv H_R+H_{DQD}
\end{equation}
is the unperturbed Hamiltonian, $E$ is its ground state energy, and $H_T$ is the tunneling Hamiltonian. Only the source reservoir is essential for the coherent indirect coupling. Thus, in the following, we consider only the source reservoir part of the tunneling Hamiltonian and omit the index $S$ for clarity. As a result, we obtain 32 terms that contribute to the kinetic exchange interaction. In such contributions, the most dominant contribution has the form
\begin{eqnarray}
&&2\sum_{|k|>k_F}\sum_{|k'|\le k_F}\frac{{t_{k'}^{(1)}}^*t_{k'}^{(2)}}{\epsilon_k-\epsilon_{k'}+i\eta}\frac{{t_{k}^{(2)}}^*t_{k}^{(1)}}{\left(\epsilon_k+\frac{U}{2} \right)^2}{d_{1\downarrow}}^{\dagger}{d_{2\uparrow}}^{\dagger}|F\rangle\nonumber\\
&&+2\sum_{|k|>k_F}\sum_{|k'|\le k_F}\frac{{t_{k'}^{(1)}}^*t_{k'}^{(2)}}{\epsilon_k-\epsilon_{k'}+i\eta}\frac{{t_{k}^{(2)}}^*t_{k}^{(1)}}{\left(\epsilon_k-\frac{U}{2} \right)^2}{d_{1\downarrow}}^{\dagger}{d_{2\uparrow}}^{\dagger}|F\rangle\nonumber\\
&&-4\sum_{|k|>k_F}\sum_{|k'|\le k_F}\frac{{t_{k'}^{(1)}}^*t_{k'}^{(2)}}{\epsilon_k-\epsilon_{k'}+i\eta}\frac{{t_{k}^{(2)}}^*t_{k}^{(1)}}{\left(\epsilon_k+\frac{U}{2} \right)\left(\epsilon_{k'}-\frac{U}{2} \right)}{d_{1\downarrow}}^{\dagger}{d_{2\uparrow}}^{\dagger}|F\rangle,\label{eff-ham1}
\end{eqnarray}
where $\eta$ is positive infinitesimal, and we focused on the particle-hole symmetric condition, namely $\epsilon_1=\epsilon_2=-V_{inter}-\frac{U}{2}$. These have one electron-hole excitation pair in the intermediate states, and this pair leads to the energy denominator of $\epsilon_k-\epsilon_{k'}$. In Eq. (\ref{eff-ham1}), we only need to consider the low energy excitation in the vicinity of the Fermi surface because of the energy denominator $\epsilon_k-\epsilon_{k'}$ and the condition $\epsilon_{k'}<\epsilon_F<\epsilon_k$, where $\epsilon_F$ is the Fermi energy. Moreover, we can neglect $\epsilon_k$ in $\epsilon_k\pm\frac{U}{2}$ since $|\epsilon_k|\ll \frac{U}{2}$. Thus, we have
\begin{eqnarray}
8\left(\frac{2}{U} \right)^2\sum_{|k|>k_F}\sum_{|k'|\le k_F}\frac{{t_{k'}^{(1)}}^*t_{k'}^{(2)}{t_{k}^{(2)}}^*t_{k}^{(1)}}{\epsilon_k-\epsilon_{k'}+i\eta}{d_{1\downarrow}}^{\dagger}{d_{2\uparrow}}^{\dagger}|F\rangle.
\end{eqnarray}
Although we have to estimate the wave number integration, according to the prescription given in Ref. \onlinecite{kubo1}, the azimuthal integration gives rise to the oscillatory behavior of the coherent indirect coupling parameter with respect to the propagation length, and the radial integration is
\begin{eqnarray}
8\left(\frac{2}{U} \right)^2\int_{-\epsilon_F}^{\epsilon_F}\frac{d\epsilon}{2\pi}f(\epsilon)\int_{-\epsilon_F}^{\epsilon_F}\frac{d\epsilon'}{2\pi}[1-f(\epsilon')]\frac{\Gamma_{12}(\epsilon)\Gamma_{21}(\epsilon')}{\epsilon-\epsilon'+i\eta}{d_{1\downarrow}}^{\dagger}{d_{2\uparrow}}^{\dagger}|F\rangle.
\end{eqnarray}
In the wide-band limit, we neglect the energy dependence of the linewidth functions, and thus we obtain
\begin{eqnarray}
H_{eff}^{\alpha}\simeq16\epsilon_F\left(\frac{\alpha\hbar\Gamma}{\pi U} \right)^2\bm{S}_1\cdot\bm{S}_2,
\end{eqnarray}
where we have neglected the spin-independent terms. Therefore, the exchange coupling constant is
\begin{eqnarray}
J_{\alpha}=16\epsilon_F\left(\frac{\alpha\hbar\Gamma}{\pi U} \right)^2.
\end{eqnarray}

\section{\label{spin-spin}Effect of coherent indirect coupling for both source and drain reservoirs on spin-spin correlation}
\begin{figure}
\includegraphics[scale=0.6]{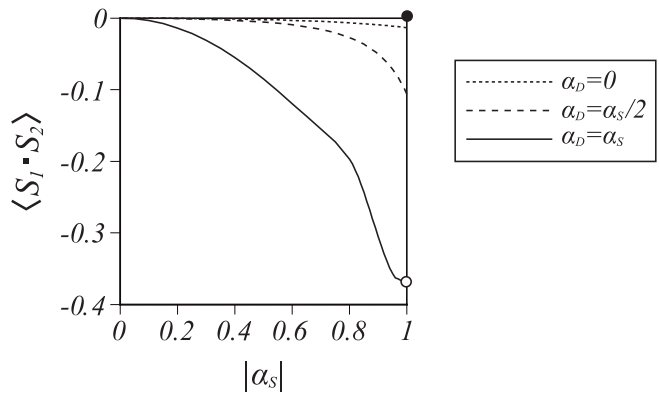}% Here is how to import Epseudospin art
\caption{\label{fig5} For various quotients between $\alpha_S$ and $\alpha_D$, $|\alpha_S|$ dependence of the spin-spin correlation function at $\epsilon_1/\hbar\Gamma=\epsilon_2/\hbar\Gamma=-6$ indicated by the green circle in Fig. \ref{fig3} (a).}
\end{figure}

In this Appendix, we show the $\alpha$ dependence of the spin-spin correlation when the coherent indirect couplings are considered for both the source and drain reservoirs as discueed in Sec. \ref{result} A. Then, we define the coherent indirect coupling parameter of the reservoir $\nu$ ($\nu\in\{S,D\}$) as $\alpha_{\nu}$. Then, in Fig. \ref{fig5}, we plot the $|\alpha_S|$ dependence of the spin-spin correlation function for various quotients between $\alpha_S$ and $\alpha_D$ at $\epsilon_1/\hbar\Gamma=\epsilon_2/\hbar\Gamma=-6$ indicated by the green circle in Fig. \ref{fig3} (a). It is clear that we have a stronger suppression of Kondo effect due to an antiferromagnetic kinetic exchange coupling induced by the coherent indirect couplings for both the source and drain reservoirs in comparison with the result shown in Fig. \ref{fig3} (c). As shown in Fig. \ref{fig5}, the spin-spin correlation vanishes at $|\alpha_S|=|\alpha_D|=1$. Under this condition, there is only a single conduction mode \cite{kubo1}. As a result, we have the single channel spin Kondo effect, and the linear conductance has a value of $2e^2/h$.

\newpage %Just because of unusual number of tables stacked at end


\begin{thebibliography}{99}
\bibitem{alpha} Here we assume that the orbital states of the two QDs are the same. In general, $|\alpha|=1$.
\end{thebibliography}
\end{document}